\documentclass{ws-p9-75x6-50}

\begin{document}

\title{The cosmic microwave background bispectrum as a test of the physics of inflation and probe of the astrophysics of the low-redshift universe}

\author{Eiichiro Komatsu$^{1,2,3}$ and David N. Spergel$^{1,4}$}

\address{$^1$ Department of Astrophysical Sciences, Princeton University,
Princeton, NJ 08544, USA\\ 
$^2$ School of Natural Sciences, Institute for Advanced Study, 
Princeton, NJ 08540, USA\\
$^3$ Astronomical Institute, T\^ohoku University, 
Aoba, Sendai 980-8578, Japan\\
$^4$ Keck Distinguished Visiting Professor,
School of Natural Sciences, Institute for Advanced Study, 
Princeton, NJ 08540, USA\\
E-mail: komatsu@astro.princeton.edu, dns@astro.princeton.edu}

\maketitle

\abstracts{ 
Why is non-Gaussianity interesting? One of generic predictions from 
inflationary scenarios is that primordial fluctuations are exactly Gaussian 
in linear order; however, the non-linearity in the inflation 
will produce weak non-Gaussianity. Thus, measuring the non-Gaussianity
in the cosmic microwave background radiation anisotropy
is a probe of the non-linear physics in the very early universe.
Since the angular three point function is zero for the Gaussian field, 
it is sensitive to the non-Gaussianity. We predict 
its harmonic transform counterpart, the angular {\it Bispectrum}, 
down to arcminutes angular scales, including the full effect of 
the radiation transfer function.
We find that even the Planck experiment cannot detect 
the primary bispectrum from the inflation, as long as the single field 
slow-roll inflation is right.
Non-linearities in the low redshift universe also produce the
non-Gaussianity. We find that secondary bispectra are detectable
by both MAP and Planck experiments. The secondary bispectra 
probe the non-linear physics of the low-redshift universe.
Although this could be a contaminant 
to the primary signal,
MAP and Planck experiments are found to be able to separate the primary
from secondary effects well. 
We present a tentative comparison of the primary bispectrum to
the published COBE 4 year bispectrum. The data put a weak constraint
on the parameter, and the constraint would become much tighter
when we use all modes available in the COBE data, and 
certainly forthcoming satellite experiments. 
As a conclusion, the bispectrum is 
a key measure to confirm or destroy the simple inflationary scenario
in non-linear order that seems quite successful in linear order.
}
\section{The CMB non-Gaussianity from inflation}
\subsection{Order-of-magnitude estimation of the non-linear coupling
  parameter $f_{NL}$}\label{sec:fNL}

We briefly consider the mechanism producing the primary CMB non-Gaussianity
from the inflation. Since the observed CMB anisotropy $\Delta T/T$ 
is thought to be generated by curvature perturbations $\Phi$,
the primary non-Gaussianity should be encapsulated in $\Phi$.
The inflation makes $\Phi$ weakly non-Gaussian.
However, even if $\Phi$ is Gaussian, $\Delta T/T$ can be
non-Gaussian. 

$\Delta T/T$ is related to $\Phi$ through the 
non-linear relation: 
\begin{equation}
 \label{eq:T-Phi}
  \frac{{\Delta T}}T\sim g_{T}\left(\Phi+f_\Phi \Phi^2\right),
\end{equation}
where $g_T$ is the radiation transfer function converting
$\Phi$ to $\Delta T/T$. $g_T=-1/3$ for the Sachs-Wolfe effect\cite{SW67}.
The second term with $f_\Phi\sim {\cal O}(1)$
is the higher order correction that arises from the 
second order gravitational perturbation theory\cite{PC96}.
Thus, it makes $\Delta T$ weakly non-Gaussian, even when
$\Phi$ is Gaussian. 
$\Phi$ is actually a weakly non-Gaussian field, and thought to be 
produced from the inflaton fluctuations $\delta\phi$ through
the relation\cite{SB9091}:
\begin{equation}
 \label{eq:Phi-inflaton}
  \Phi\sim m_{pl}^{-1}g_{\Phi}
           \left(\delta\phi+m_{pl}^{-1}f_{\delta\phi}\delta\phi^2\right).
\end{equation}
$g_{\Phi}\sim {\cal O}(10)$, and $f_{\delta\phi}\sim 
{\cal O}(10^{-1})$ 
for a class of slow-roll inflationary scenarios 
with the single inflaton field.
If the dynamics of $\delta\phi$ field is simple enough to keep itself 
Gaussian throughout the evolution, then we can stop our 
consideration here. 
However, in generic, 
the non-trivial form of the interaction term 
in the equation of motion\cite{FRS93},
or the non-linear coupling between the coarse-grained classical fluctuations 
and the fine scale quantum fluctuations\cite{GLMM94,Starobinsky86},
make $\delta\phi$ weakly non-Gaussian, even if it is originally 
Gaussian field. The result is simply written as
\begin{equation}
 \label{eq:inflaton-eta}
  \delta\phi\sim g_{\delta\phi}
                 \left(\eta+m_{pl}^{-1}f_{\eta}\eta^2\right),
\end{equation}
where $\eta$ is Gaussian, $g_{\delta\phi}\sim 1$, and
$f_{\eta}\sim {\cal O}(10^{-1})$.

It is then useful to rewritten 
equations (\ref{eq:T-Phi})--(\ref{eq:inflaton-eta}) as
\begin{equation}
 \label{eq:T-Phi_L}
  \frac{{\Delta T}}T\sim 
  g_{T}\left[\Phi_L
            +\left(f_\Phi
                  +g_\Phi^{-1}f_{\delta\phi}
                  +g_\Phi^{-1}g_{\delta\phi}^{-1}f_{\eta}
             \right)\Phi_L^2\right],
\end{equation}
where $\Phi_L\equiv g_\Phi g_{\delta\phi}\eta \sim 10\eta$ 
is the Gaussian part of curvature perturbations.
More specifically, we define the {\it non-linear coupling parameter},
$f_{NL}$, as
\begin{equation}
 \label{eq:Phi}
  \Phi({\mathbf x})
  = \Phi_L({\mathbf x})
   +f_{NL}\left(
          \Phi^2_{L}({\mathbf x})-\left<\Phi^2_{L}({\mathbf x})\right>
          \right),
\end{equation}
where the angular bracket denotes the statistical ensemble average.
Note that $f_{NL}\sim f_\Phi 
+g_\Phi^{-1}f_{\delta\phi}
+g_\Phi^{-1}g_{\delta\phi}^{-1}f_{\eta}$, and 
the first term $\sim {\cal O}(1)$ 
from the second order gravity effect is dominant 
compared to other two terms from slow-roll inflation $\sim {\cal O}(10^{-2})$.
$f_{NL}$ corresponds to $\alpha_\Phi$ in Verde et al.\cite{VWHK00}. 

\subsection{The reduced angular bispectrum}

We use the angular bispectrum, harmonic transform of the 
angular three point function, as a measure of the CMB non-Gaussianity.

$\Delta T(\hat{\mathbf n})/T$ in the sky is expanded into the 
spherical harmonics: 
$\Delta T(\hat{\mathbf n})/T=\sum_{lm}a_{lm}Y_{lm}(\hat{\mathbf n})$.
$a_{lm}$ is then written in terms of $\Phi$-field,
\begin{equation}
  \label{eq:almphi}
  a_{lm}=4\pi(-i)^l
  \int\frac{d^3{\mathbf k}}{(2\pi)^3}\Phi({\mathbf k})g_{Tl}(k)
  Y_{lm}^*(\hat{\mathbf k}).
\end{equation}
$\Phi({\mathbf k})$ is weakly non-Gaussian field given by the 
Fourier transform of equation (\ref{eq:Phi}).
Instead of assuming the Sachs-Wolfe's $-1/3$ law for $g_{Tl}(k)$, 
we use the full $g_{Tl}(k)$ down to small scales by solving
the Boltzmann transport equation with the CMBFAST code\cite{SZ96}.
We then define the {\it reduced} angular bispectrum
$b_{l_1l_2l_3}$ as\cite{KS00}
\begin{eqnarray}
  \nonumber
  b_{l_1l_2l_3}&\equiv&
  \sqrt{\frac{4\pi}{(2l_1+1)(2l_2+1)(2l_3+1)}}
  \left(
  \begin{array}{ccc}
  l_1&l_2&l_3\\
  0&0&0
  \end{array}
  \right)^{-1}\\
  & &\times
  \label{eq:reducedb}
  \sum_{m_1m_2m_3}
  \left(
  \begin{array}{ccc}
  l_1&l_2&l_3\\
  m_1&m_2&m_3
  \end{array}
  \right)
  \left<a_{l_1m_1}a_{l_2m_2}a_{l_3m_3}\right>,
\end{eqnarray}
where the matrix is the Wigner-3$j$ symbol.
$b_{l_1l_2l_3}$ is found to be more useful to describe the physical 
properties of the angular bispectrum than the conventional 
angular averaged bispectrum,
$B_{l_1l_2l_3}\equiv 
\sum_{m_1m_2m_3}
  \left(
  \begin{array}{ccc}
  l_1&l_2&l_3\\
  m_1&m_2&m_3
  \end{array}
  \right)
  \left<a_{l_1m_1}a_{l_2m_2}a_{l_3m_3}\right>$.
Note that Magueijo\cite{Magueijo00} defines a similar quantity, 
$\hat{B}_{l_1l_2l_3}=b_{l_1l_2l_3}/\sqrt{4\pi}$.

We give the explicit form of the primary reduced bispectrum 
in the reference\cite{KS00}. 
Figure~1 plots the equilateral case ($l_1=l_2=l_3\equiv l$)
reduced bispectrum. The acoustic oscillation in $g_{Tl}(k)$
appears in $b_{lll}$. For the scale invariant $\Phi_L$-field power
spectrum, 
\begin{equation}
 \label{eq:order_primary}
  b^{primary}_{lll}\sim 10^{-17}f_{NL}\times l^{-4},
\end{equation}
apart from the oscillation and the Silk damping.

\begin{figure}[t]
\begin{center}
\epsfxsize=25pc 
\epsfbox{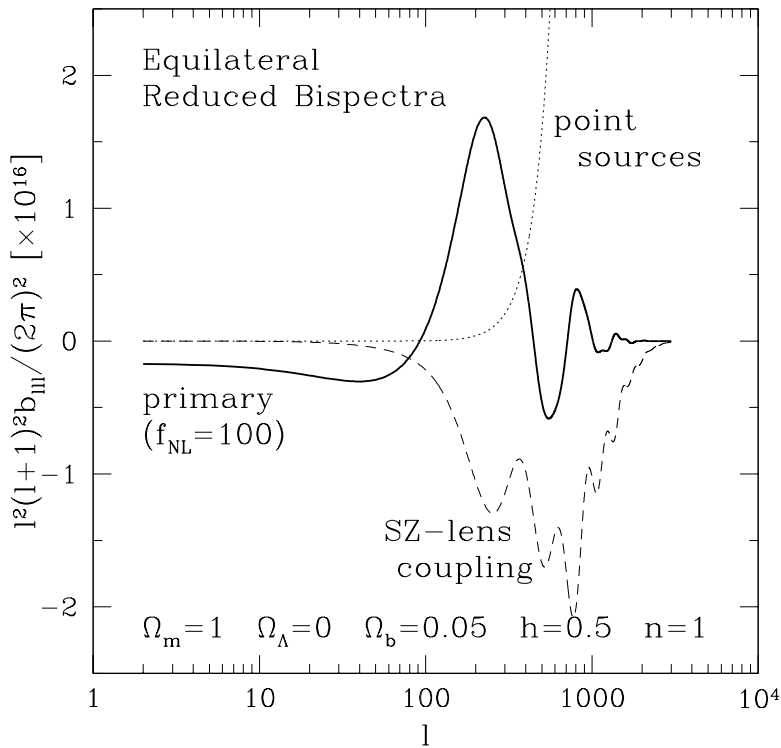} 
\end{center}
\caption{The equilateral configuration of the reduced bispectra,
$\left[l^2(l+1)^2b_{lll}/(2\pi)^2\right]\times 10^{16}$. 
The weight is motivated by the behavior 
of the primary component, $b_{lll}^{primary}\propto l^{-4}$.
Solid line shows the primary bispectrum from inflation with $f_{NL}=100$
(see eq.(\ref{eq:Phi})).
Dashed line shows the Sunyaev-Zel'dovich effect and weak lensing
 coupling in the Rayleigh--Jeans limit,
while dotted line shows the extragalactic radio and infrared point sources.  
}
\end{figure}

\section{The CMB non-Gaussianity from low-redshift universe}

Secondary bispectra from the coupling between 
the Sunyaev-Zel'dovich (SZ) effect and 
the weak lensing effect\cite{GS99,CH00}, and the extragalactic 
radio and infrared sources\cite{KS00,RSH00} are 
found to be marginally detectable by MAP, while 
Planck will detect them significantly.

We give the explicit form of the SZ--lensing coupling bispectrum in 
the reference\cite{KS00}. It is roughly
\begin{equation}
 \label{eq:order_sz-lens}
  b^{sz-lens}_{lll}\sim 10^{-19}j_\nu\overline{T}_{\rho0}b_{gas}\times l^{-3},
\end{equation}
up to $l\sim 1000$. $\overline{T}_{\rho0}\sim 0.3$ is the 
present day density weighted 
mean temperature in units of 1 keV, while 
$b_{gas}\sim 6$ is the linear gas pressure bias\cite{RKSP00}.
$j_\nu$ is the spectral function of the SZ effect, and equals 
$-2$ in the Rayleigh--Jeans regime\cite{ZS69}.

The CMB bispectrum from extragalactic point sources is easy to 
estimate, as long as they are assumed to be Poisson distributed.
The reduced bispectrum is just constant in that case, i.e., 
$b_{l_1l_2l_3}^{ps}=b^{ps}={\rm constant}$.
Detailed arguments are summarized in the reference\cite{KS00}. The 
results are
\begin{eqnarray}
  \label{eq:bl90ghz}
  b^{ps}(90~{\rm GHz},<2~{\rm Jy})&\sim& 2\times 10^{-25},\\
  \label{eq:bl217ghz}
  b^{ps}(217~{\rm GHz},<0.2~{\rm Jy})&\sim& 5\times 10^{-28}.
\end{eqnarray}
Equations (\ref{eq:bl90ghz}) and (\ref{eq:bl217ghz})
correspond to MAP and Planck experiments, respectively,
after subtracting $>5\sigma$ sources from the map.

Figure~1 plots $b_{lll}^{sz-lens}$ in the Rayleigh--Jeans limit 
and $b^{ps}$. Clearly, secondary sources dominate the primary signal
as going to smaller angular scales, even if $f_{NL}=100$ is taken.
This is opposite to the CMB power spectrum that is dominated by
either the primary CMB or the instrumental noise at all $l$ in the
COBE, MAP, and Planck experiments.
Thus, we need to separate the primary bispectrum from secondary 
bispectra.
Fortunately, the shapes of bispectra are fairly different, 
so that we can discriminate among them based on shapes.

\section{Can we detect/separate the CMB bispectra?}

Since the CMB bispectrum is very weak compared to the Gaussian noise
from CMB itself, it is totally hopeless to measure $b_{l_1l_2l_3}$
{\it per mode}. Instead, we fit the observed data by
theoretical bispectra, and then constrain parameters. For this
purpose, we minimize $\chi^2$ given by
\begin{equation}
  \label{eq:chisq}
  \chi^2(A_i)
  \equiv 
  \sum_{2\le l_1\le l_2\le l_3}
  \frac{\left(B_{l_1l_2l_3}^{obs}
	     -\sum_i A_i B^{(i)}_{l_1l_2l_3}\right)^2}
  {\sigma^2_{l_1l_2l_3}},
\end{equation}
where $i$ denotes a component such as
the primary, the SZ and lensing effects, extragalactic sources, and so on. 
In case that the non-Gaussianity is small, the cosmic variance
of the bispectrum $\sigma^2_{l_1l_2l_3}$
is given by the six-point function of $a_{lm}$\cite{SG99,GM00}.
$\sigma^2_{l_1l_2l_3}$ is dominated by either 
the primary CMB or the instrumental 
noise, while the secondary sources are subdominant as long as
COBE, MAP, and Planck experiments are considered.
Taking $\partial\chi^2/\partial A_i=0$, we obtain the Fisher matrix,
\begin{equation}
  \label{eq:fis}
  F_{ij}\equiv 
  \sum_{2\le l_1\le l_2\le l_3}
  \frac{B_{l_1l_2l_3}^{(i)}B_{l_1l_2l_3}^{(j)}}{\sigma_{l_1l_2l_3}^2}.
\end{equation}
Since the covariance matrix of $A_i$ is $F_{ij}^{-1}$,
we define the signal-to-noise ratio 
$(S/N)_i\equiv 1/\sqrt{F_{ii}^{-1}}$ for 
a component $i$, and
the signal degradation parameter 
$d_i\equiv F_{ii}F_{ii}^{-1}$ due to a contamination from
correlated components.
$d_i$ is defined so as $d_i=1$ for perfectly separated component,
while $d_i>1$ for poorly separated one.
Table~1 and 2 summarize $(S/N)_i$ and $d_i$, respectively,
for COBE, MAP, and Planck experiments.
In order to obtain $(S/N)_{primary}>1$, 
$f_{NL}>600$, 20, and 5 are needed for COBE, MAP, and Planck
experiments.
As derived in section~\ref{sec:fNL}, the single field slow-roll 
inflation generally predicts $f_{NL}\sim {\cal O}(10^{-2})$, while
the second order gravity yields $f_{NL}\sim {\cal O}(1)$.
Therefore, either way, the detection is not expected for those
experiments.

\begin{table}[t]
\caption{The signal-to-noise ratio for the detection of
bispectra.
$f_{NL}$ is the non-linear coupling parameter (Eq.~(\ref{eq:Phi})).
$\overline{T}_{\rho0}\sim 0.3$ is the present day density weighted 
mean temperature in units of 1 keV, while 
$b_{gas}\sim 6$ is the linear gas pressure bias.
$j_\nu$ is the spectral function of the Sunyaev-Zel'dovich effect that 
has $-2$ in the Rayleigh--Jeans regime.
$b^{ps}_{25}\equiv b^{ps}/10^{-25}$, and
$b^{ps}_{27}\equiv b^{ps}/10^{-27}$, where $b^{ps}$ is the reduced 
Poisson bispectrum.
}
\begin{center}
\footnotesize
\begin{tabular}{c|ccc}
 \hline
       & primary & SZ--lensing & point sources \\
  \hline 
  COBE & $1.7\times 10^{-3}~f_{NL}$ 
       & $1.8\times 10^{-4}~\left|j_\nu\right| \overline{T}_{\rho0}b_{gas}$
       & $5.7\times 10^{-7}~b_{25}^{ps}$ \\
  MAP  & $5.8\times 10^{-2}~f_{NL}$ 
       & $0.34~\left|j_\nu\right| \overline{T}_{\rho0}b_{gas}$
       & $2.2~b_{25}^{ps}$ \\
  Planck & $0.19~f_{NL}$ 
         & $6.2~\left|j_\nu\right| \overline{T}_{\rho0}b_{gas}$
         & $52~b_{27}^{ps}$ \\
 \hline
\end{tabular}
\end{center}
\end{table}

\begin{table}[t]
\caption{The signal degradation parameters.
$d_i$ is defined so as $d_i=1$ for perfectly separated component,
while $d_i>1$ for poorly separated one.
}
\begin{center}
\footnotesize
\begin{tabular}{c|ccc}
 \hline
       & primary & SZ--lensing & point sources \\
  \hline 
  COBE & 1.46 & 3.89 & 3.45 \\ 
  MAP  & 1.01 & 1.16 & 1.14 \\
  Planck & 1.00 & 1.00 & 1.00 \\
 \hline
\end{tabular}
\end{center}
\end{table}

$(S/N)_{sz-lens}\sim 1$ and $(S/N)_{ps}\sim 4$ are
expected for MAP,
and $(S/N)_{sz-lens}\sim 20$ and $(S/N)_{ps}\sim 30$ for 
Planck. Thus, the leading contribution to the CMB bispectrum
would be from extragalactic point sources.
Then, the SZ--lensing contribution would be measured by
Planck, yielding a constraint on $\overline{T}_{\rho0}b_{gas}$.
This quantity measures the physical state of intergalactic gas 
pressure that is uncertain at present.

\section{Tentative comparison to the COBE 4 year bispectrum}

Ferreira, Magueijo, and G\'orski\cite{FMG98} claimed 
a significant detection of the equilateral bispectrum at $l=16$
in the publicly available COBE 4 year data.
Banday, Zaroubi, and G\'orski\cite{BZG00} found that
their detection would be ascribed to the systematics associated
with the eclipse of COBE satellite.
This partly demonstrates that the bispectrum is sensitive to 
the non-linear effect even including the systematics in data.

Magueijo\cite{Magueijo00} tabulates a particular configuration
of the COBE bispectrum without eclipse data.
He tabulates 8 modes of 
\begin{equation}
 \label{eq:Jl3}
  J_l^3\equiv 
  \frac{b_{l-1ll+1}}{\sqrt{4\pi C_{l-1}C_lC_{l+1}}},
\end{equation}
for $l=4-18$.
Figure~2 compares our primary bispectrum to his data.
There is no significant detection of $J_l^3$ per mode,
and then $f_{NL}$ is constrained as $f_{NL}<10^4$.
It should be stressed that only 8 modes were used to constrain $f_{NL}$
here. Since the total number of modes available in the full COBE bispectrum
is 466 up to $l=20$, the constraint on $f_{NL}$ becomes much tighter than
$<10^4$, say, $1\sigma\sim 600$ as suggested by our estimation (table~1).
MAP and Planck will certainly put a very stringent constraint on
$f_{NL}$, and thus test the inflationary scenario in non-linear order.

\begin{figure}[t]
\begin{center}
\epsfxsize=25pc 
\epsfbox{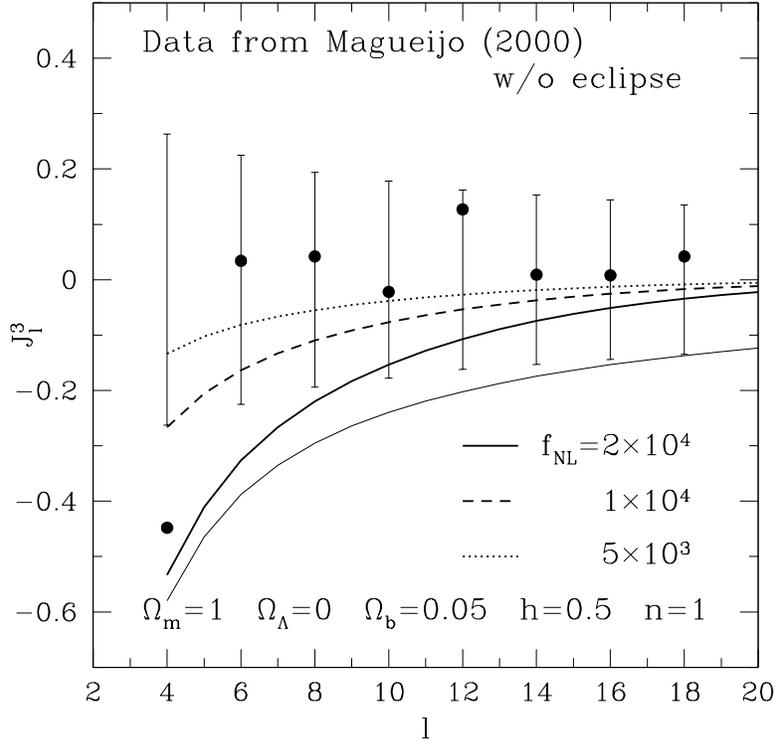} 
\end{center}
\caption{The comparison of theoretical primary bispectrum to the 
COBE 4 year data, $J_l^3$ (Eq.(\ref{eq:Jl3})), 
defined by Magueijo (2000).
Filled circles are measured $J_l^3$, while
thick solid, dashed, and dotted lines correspond to the primary
bispectrum with $f_{NL}=2\times 10^4$, $10^4$, and $5\times 10^3$,
respectively. Error bars are rms scatters from the 
Gaussian Monte Carlo realizations. 
Note that the beam-smearing effect has been taken into account.
The thin solid line shows the un-smoothed case for 
$f_{NL}=2\times 10^4$.}
\end{figure}

\section{Discussion}

What if the primary bispectrum is detected? In other words, 
what if $f_{NL}\gg 1$ is measured?
One might suspect the systematics, as already demonstrated in the COBE
data, or foreground secondary sources. 
Since we fit the theory to the data, the 
{\it goodness-of-fit} value of $\chi^2$
at the best-fit $f_{NL}$ would tell us whether detected signal is 
really primary or not.
If it passes the $\chi^2$ test, then the most straightforward consequence
is that --- the single field slow-roll inflation is not right.

\section*{Acknowledgments}

We would like to thank Licia Verde for invaluable discussion.
E. K. acknowledges a fellowship
from the Japan Society for the Promotion of Science.
D. N. S. is partially supported by the MAP/MIDEX program.



\begin{thebibliography}{99}

\bibitem{SW67}
R. K. Sachs and A. M. Wolfe,
{\it Astrophys. J.} {\bf 147}, 73 (1967).

\bibitem{PC96}
T. Pyne and S. M. Carroll,
\Journal{\PRD}{53}{2920}{1996}.

\bibitem{SB9091} 
D. S. Salopek and J. R. Bond,
\Journal{\PRD}{42}{3936}{1990};
{\it ibid.} {\bf 43}, 1005 (1991).

\bibitem{GLMM94} 
A. Gangui, F. Lucchin, S. Matarrese and S. Mollerach,
{\it Astrophys. J.} {\bf 430}, 447 (1994).

\bibitem{FRS93}
T. Falk, R. Rangarajan and M. Srednicki, 
{\it Astrophys. J. Lett.} {\bf 403}, L1 (1993).

\bibitem{Starobinsky86}
A. A. Starobinsky,
in {\it Field Theory, Quantum Gravity, and Strings},
edited by H. T. de Vega and N. Sanchez, 
Lecture Notes in Physics, Vol. 246 (Springer-Verlag, Berlin, 1986), p.107.

\bibitem{VWHK00}
L. Verde, L. Wang, A. F. Heavens and M. Kamionkowski,
Mon. Not. R. Astron. Soc. {\bf 313}, 141 (2000).

\bibitem{SZ96} 
U. Seljak and M. Zaldarriaga,
{\it Astrophys. J.} {\bf 469}, 437 (1996).

\bibitem{KS00}
E. Komatsu and D. N. Spergel,
{\it Phys. Rev. D} (submitted); {\it preprint}, astro-ph/0005046.

\bibitem{Magueijo00}
J. Magueijo,
{\it Astrophys. J. Lett.} {\bf 528}, L57 (2000).

\bibitem{GS99} 
D. M. Goldberg and D. N. Spergel,
\Journal{\PRD}{59}{103002}{1999}.

\bibitem{CH00}
A. Cooray and W. Hu,
{\it Astrophys. J.} {\bf 534}, 533 (2000).

\bibitem{RSH00}
A. Refregier, D. N. Spergel and T. Herbig,
{\it Astrophys. J.} {\bf 531}, 31 (2000).

\bibitem{RKSP00}
A. Refregier, E. Komatsu, D. N. Spergel and U.-L. Pen,
\Journal{\PRD}{61}{123001}{2000}.

\bibitem{ZS69}
Ya. B. Zel'dovich and R. A. Sunyaev,
{\it Astrophys. Space. Sci.} {\bf 4}, 301 (1969).

\bibitem{SG99} 
D. N. Spergel and D. M. Goldberg,
\Journal{\PRD}{59}{103001}{1999}.

\bibitem{GM00}
A. Gangui and J. Martin,
Mon. Not. R. Astron. Soc., {\bf 313}, 323 (2000);
\Journal{\PRD}{62}{103004}{2000}.

\bibitem{FMG98} 
P. G. Ferreira, J. Magueijo and K. M. G\'orski,
{\it Astrophys. J. Lett.} {\bf 503}, L1 (1998).

\bibitem{BZG00}
A. J. Banday, S. Zaroubi and K. M. G\'orski,
{\it Astrophys. J.} {\bf 533}, 575 (2000). 

\end{thebibliography}
\end{document}